\documentclass[seceq]{ptptex}

\usepackage{graphicx}

\title{
Pattern of Light Scalar Mesons
}

\author{
Keh-Fei \textsc{Liu}\,\footnote{ e-mail address:
liu@pa.uky.edu}
}

\inst{
Dept. of Physics and Astronomy, University of Kentucky, Lexington,
KY 40506 USA
}

\abst{
Combining the recent lattice calculation of $a_0(1450)$ and
$\sigma(600)$ mesons with the overlap fermion in the chiral regime
with the pion mass less than $300\,{\rm MeV}$, the quenched lattice calculation
of the scalar glueball, and the phenomenological study of the mixing of 
isoscalar scalar mesons $f_0(1710)$, $f_0(1500)$, $f_0(1370)$ through their decays, 
a simple pattern for the light scalar mesons begins to emerge. Below 1 GeV,
the scalar mesons form a nonet of tetraquark mesoniums. Above 1
GeV, the nonent $q\bar{q}$ mesons are made of an octet with largely
unbroken $SU(3)$ symmetry and a fairly good singlet which is $f_0(1370)$. 
$f_0(1710)$ is identified as an almost pure scalar glueball with a $\sim 10\%$
mixture of $q\bar{q}$.}

\begin{document}

\maketitle

\section{Introduction}   \label{intro}

\begin{figure}[tbh]
  \centering
  \includegraphics[width=10.0cm]{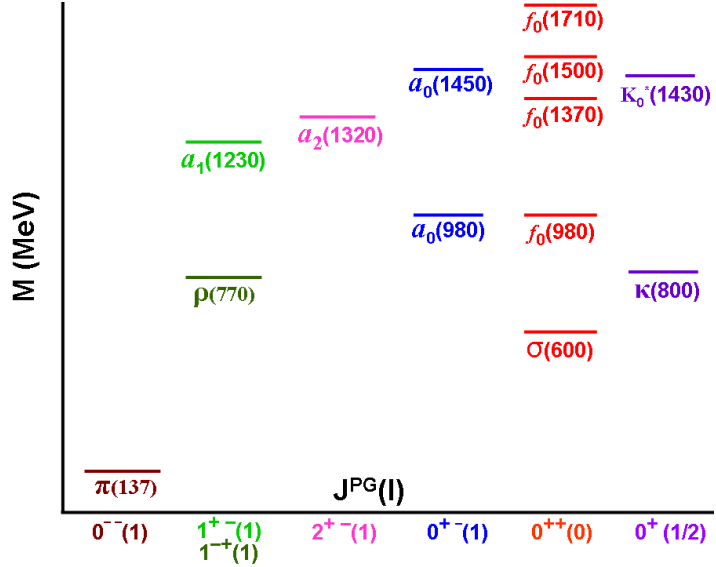}
  \caption{ Spectrum of scalar mesons together with
 $\pi$, $\rho, a_1$ and $a_2$ mesons.}
 \label{fig:scalar}
\end{figure}

    In light meson spectroscopy, the pseudoscalar, vector, axial, and tensor
sectors are reasonably well known in terms of their $SU(3)$
classification and quark contents. The scalar sector, on the other
hand, is poorly understood in this regard. First of all, there are
too many of them. There are 19 states which are more than twice
the usual $q\bar{q}$ nonet as in other sectors. We show in
Fig.~\ref{fig:scalar} the known scalar mesons which include
$\sigma(600), \kappa(800)$, and $f_0(1710)$ which are better
established experimentally nowadays~\cite{pdt06,ait01}. There are
several puzzling characteristics which have been observed over
the years. The first question one might raise is the whereabout of the
$q\bar{q}$ $a_0$, the ${}^3P_0$ partner of
$a_1(1260)\,({}^3P_1)$ and $a_2(1320)\,({}^3P_2)$ according to the
quark model classification. From the order of spin-orbit splitting
of the P-wave $q\bar{q}$ spectrum, it seems natural to identify it with
$a_0(980)$. However, there are a host of difficulties in such an
assignment:

\begin{itemize}
\item
  In this case, the member of the octet $K_0^*$ (e.g. $s\bar{u}$ with one strange quark) is
expected to lie $\sim 100$ MeV above, which would place it around 1100 MeV. But there is no
state there, it would be $\sim 300$ MeV below $K_0^*(1430)$ and $\sim 300$ MeV above
$\kappa(800)$.

\item
 The widths of $a_0(980)$ and $f_0(980)$ are substantially smaller than those of
   $a_0(1450)$ and $f_0(1370)$. In particular, they are much smaller than that of $\kappa(800)$
   which should be a nonet partner of $a_0(980)$ and $f_0(980)$.

\item
The $\gamma\gamma$ widths of $a_0(980)$ and $f_0(980)$ are much smaller than expected
   of a $q\bar{q}$ state~\cite{bar85}.

\item
 It is hard to understand why $a_0(980)$ and $f_0(980)$ are practically degenerate.
   The experimental data on $D_s^+ \rightarrow f_0(980)\pi^+$  and $\phi \rightarrow f_0(980)
   \gamma$ imply copious $f0(980)$ production via its $s\bar{s}$ component. Yet, 
   there cannot be $s\bar{s}$ in $a_0(980)$ since it is an $I=1$ state.

\item The radiative decay $\phi\to a_0(980)\gamma$, which cannot
    proceed if $a_0(980)$ is a $q\bar q$ state, can be nicely
    described in the kaon loop mechanism~\cite{Schechter06}. This
    suggests a considerable admixture of the $K\overline{K}$ component which is in contradiction
    with assigning $a_0(980)$ as the $q\bar{q}$ meson.

\end{itemize}

   Some of the above difficulties can be reconciled if $a_0(980)$ and
$f_0(980)$ are part of the nonet of four-quark (two-quarks
and two-antiquarks) mesons which was first proposed by Jaffe based on
the MIT bag model calculation~\cite{jaf77}. The ensuing potential model
studies of these four-quark mesons are also carried out~\cite{lw81,wi82}
and it was suggested that $a_0(980)$ and $f_0(980)$ are the
$K\overline{K}$ molecular states~\cite{wi82}. We shall refer them
generically as tetraquark mesoniums, not to be concerned with their
possible clustering structure. With the four quark content, it is relatively
easy to understand the degeneracy of $a_0(980)$ and $f_0(980)$
and their narrow widths. Since they have the 
quark content $u(d)\bar{u}(\bar{d})s\bar{s}$ and sit at  
$K\bar{K}$ threshold, they do not have much phase space to decay to
$K\bar{K}$ ($a_0(980)$ decay to $\eta\pi$ is suppressed by having to go
through the $s\bar{s}$ in $\eta$); whereas, $\sigma(600)$ and $\kappa(800)$ are 
relatively far above the respective $\pi\pi$ and $\pi K$ thresholds and hence 
have much larger widths. 

Recent experimental finding of $\sigma(600)$ in $D^+\rightarrow
\pi^+\pi^-\pi^+$~\cite{ait01} and $J/\Psi\rightarrow \omega
\pi^+\pi^-$~\cite{ait01} and the dispersion analysis of $\pi\pi$ scattering
with the Roy equation~\cite{ccl06} which found a resonance at
$441^{+16}_{-8}$ MeV with a width of $544^{+18}_{-25}$ MeV 
have helped establish the existence of the broad $\sigma$ resonance. 
Besides the low-lying
scalar mesons, other candidates for tetraquark mesoniums include
those vector mesons pairs produced in $\gamma\gamma$
reactions~\cite{ll82} and hadronic productions~\cite{ll83} and the
recently discovered charmed narrow resonances~\cite{swa06}.

Given that the spectrum below 1 GeV is better understood,
many questions about classification of scalar mesons above 1 GeV are
still outstanding. For example:

\begin{itemize}
\item

 The $K_0^*(1430)$, which is a $q\bar{q}$ state in all the models~\cite{pdt06},
lies higher than the axial-vector mesons $K_1(1270)$ and $K_1(1400)$. This is a 
situation which parallels the case of non-strange mesons where $a_0(1450)$ is higher 
than $a_1(1260)$ and $a_2(1320)$ and is contrary to the conventional wisdom in
the quark model as far as the order of spin-orbit splitting is concerned.

\item

 It is not clear why $K_0^*(1430)$, having one strange quark, is almost degenerate
with $a_0(1450)$, assuming the later is $(u\bar{u}-d\bar{d})/\sqrt{2}$. This is
in contrast with all the other meson sectors. 

\item

 In the $I=0$ channel, there are three states -- $f_0(1370), f_0(1500)$ and
$f_0(1710)$ and they are expected to be $(u\bar{u}+d\bar{d})/\sqrt{2}$, $s\bar{s}$ and
glueball. Which is which? Is the mixing more like that of the
pseudoscalar sector where there is substantial mixing between $(u\bar{u}+d\bar{d})/\sqrt{2}$
and $s\bar{s}$, or those of the vector and tensor sectors where the mixing are nearly ideal
between the octet and the singlet? 

\end{itemize}

\vspace{0.5cm}

    In the following, we shall use a recent lattice calculation to verify the
existence of $\sigma(600)$ as a tetraquark mesonium to help establish the classification
of the low-lying scalars below 1 GeV. We will also use lattice calculations of
$a_0(1450)$, $K_0^*(14300)$ and glueball together with the analysis of various decays
to discern the mixing among $f_0(1370), f_0(1500)$ and $f_0(1710)$. 
Based on these, a simple patter of scalar mesons is beginning to surface as will be described
in the subsequent sections.

\section{Lattice Calculation}    \label{lattice}

\begin{figure}[tbh]
\vspace*{-0.1in}
  \centerline{%
    \includegraphics[width=11.0cm]{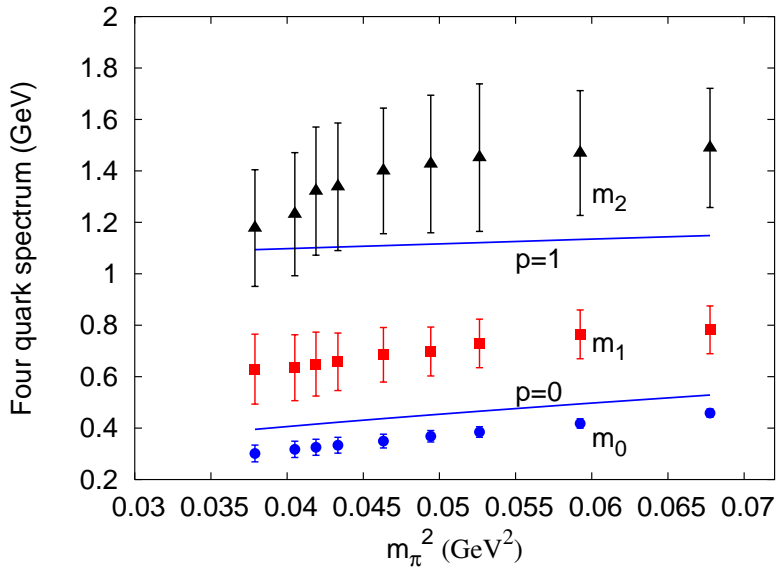}
      }
  \caption{\label{tetra}
    The lowest three states from the scalar tetraquark correlator as a function of
    $m_{\pi}^2$ for $m_{\pi}$ from 182 MeV to 250 MeV on the $12^3 \times 28$ lattice.
    The solid lines indicate the energies of the two lowest non-interacting pions in
    S-wave with lattice momenta $p_0=0$ and $p_1 = 2\pi/La$.
  }
\end{figure}

  Although there are MIT bag model~\cite{jaf77} and potential model calculations~\cite{lw81,wi82}
of tetraqurk mesoniums, lattice QCD is perhaps the most desirable theoretical tool to 
adjudicate whether these four-quark states exist and if the low-lying scalars are
indeed the predicted tetraqurk mesoniums. To begin with, we note that a resonance can be 
viewed as a mixture of a bound state and the continuum of scattering states. To establish 
the existence of a resonance on the Euclidean lattice, one can utilize the 
volume effect of a finite box where all the eigenstates are discrete (e.g. with a periodic 
boundary condition, the available momenta are $p_L= n \frac{2\pi}{La}, n =0, 
\pm 1, \pm 2 ...$) and check if there exists a bound state which is separated
from the discrete scattering states. In the context of the existence of $\sigma(600)$, 
one needs to first work in the chiral region where $m_{\pi} < 300$ MeV in recognition of 
the fact that 
the occurrence of $\sigma$ is on the basis of `current algebra, spontaneous
symmetry breakdown, and unitarity'~\cite{ccl06}. Secondly, one
needs to identify both the tetraquark mesonium and the collateral
$\pi\pi$ scattering states. Thirdly, it is necessary to work on a lattice
where the scattering states and the bound state are well separated
(e.g. further apart than half of the `would be' resonance width) in order to discern the 
nature of these states separately to make sure that $\sigma$ is indeed a one-particle
state and not a two-particle scattering state. To this end, a recent lattice QCD calculation 
was carried out on $12^3 \times 32$ and $16^3 \times 32$ lattices with $a= 0.2$ fm and
300 configurations to 
examine the spectrum with the $\overline{\Psi}\gamma_5\Psi\overline{\Psi}\gamma_5\Psi$ 
type of four-quark interpolation operators. Although a quenched calculation, it incorporates 
the chiral fermion (overlap fermion) in the chiral region with the pion mass as low as 
$182$ MeV. Results on the $12^3 \times 28$ lattice are presented in Fig.~\ref{tetra} 
as a function of $m_{\pi}^2$ for the pion mass range from 182 MeV to 250 MeV.  
The lowest state is about $100(30)$ MeV below the $\pi\pi$ threshold which is indicated
by the solid line. This is the lowest interacting state of two pions  which is attractive 
in the $I =0$ channel and is reasonably well described by the quenched chiral perturbation
theory of $\pi\pi$ scattering~\cite{bg96}. The third state is above the non-interacting 
$\pi\pi$ scattering state with each pion having one unit of lattice momentum 
(i.e. $p_1 = 2\pi/La$) and is supposed to include the higher excited states which are not fitted. 
The interesting thing is that there is an extra state
at $\sim 550$ MeV which falls in between the two states. To discern the nature of this state, we 
studied the volume dependence of the spectral weight $W_i$ from the fitting function
$\sum_i W_i e^{-E_i t}$ of the tetraquark correlator. The details are given in
Ref.~\cite{mac06}. To summarize the results, we found that, by examining the characteristic
3-volume dependence of the spectral weight, the state at  $\sim 550$ MeV is
a one-particle state, while the lowest state is a two-particle state which is
consistent with the quenched chiral perturbation prediction of the interacting $I=0$
$\pi\pi$ scattering state~\cite{bg96}.

This verifies that the tetraquark mesonium
exists and the lattice calculation, which gives a mass of $540\pm 170$ MeV at the
chiral limit, suggests that $\sigma(600)$ is such a state. 
However, one important question remains. 
Experimentally, $\sigma$ is a very broad resonance with a width
of $544$ MeV~\cite{ccl06}. How does one find its width on the lattice?
After finding $\sigma(600)$ which is separated from the $\pi\pi$ 
scattering states on the present lattice, one can increase the box which will lower the energies 
of the scattering state above it. When it is
lowered to within the range of the "width", it mixes with the
bound state and avoids level crossing. From the energy of the mixed
state one can deduce the scattering phase shift from L\"{u}scher's
formula~\cite{lus86}. This is valid for elastic scattering irrespective
how broad the resonance is. This is studied in detail in a spin
model~\cite{rg95} which illustrates how the scattering state mixes
with the bound state and gives rise to the phase shift as the
volume is increased. In a sense, by varying the lattice volume,
hence the momentum, one can use a scattering state to mix with the
bound state and scan the energy range to obtain the phase shift and
therefore the width of the resonance. The information of the width
can also be obtained by determining how far apart in energy the
scattering and bound state start to avoid the level crossing.

\begin{figure} [tbh]
  \centerline{%
    \includegraphics[width=11.0cm]{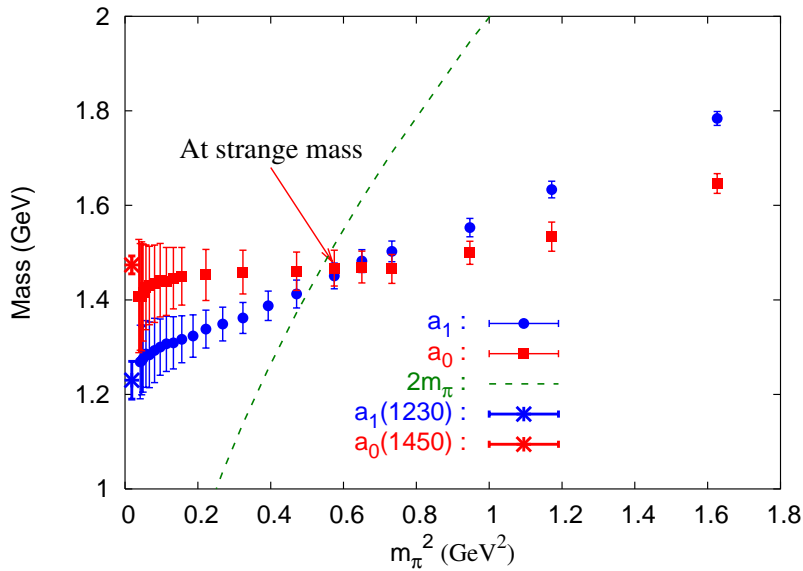}
 }
  \caption{\label{a0_a1} Masses of $a_0$, $a_1$ and two pions (dashed line) 
are plotted as a function of $m_{\pi}^2$. 
  }
\label{W_g}
\end{figure}

  To calculate $a_0$ on the same lattices, the two-quark interpolation field 
$\overline{\Psi}\Psi$ was used. We plot its mass as a function of the
corresponding $m_{\pi}^2$ in Fig.~\ref{a0_a1} together with that
of $a_1$ for comparison. We see that above
the strange quark mass, $a_1$ lies higher than $a_0$ as expected
from the quark model for heavy quarks. However, when the quark mass
is smaller than that of the strange, $a_0$ levels off, in contrast
to the $a_1$ case and those of other hadrons that have been calculated on the
lattice. This confirms the trend that has been observed in earlier
lattice calculations with higher quark masses in quenched approximation~\cite{lw00,bde02} as
well as with dynamical fermions~\cite{po03}. The
chirally extrapolated mass $a_0 = 1.42 \pm 0.13$ GeV suggests that
the meson $a_0(1450)$ is a $q\bar{q}$ state. By virtue of the fact that we
do not see $a_0(980)$, its $q\bar{q}$ content is estimated to be
two orders of magnitude smaller than that of $a_0(1450)$~\cite{mac06}.
The $K_0^{*}(1430)$ mass at $1.41 \pm 0.12$ GeV is calculated with the
strange mass fixed to reproduce the $\phi$ mass and the $u/d$ extrapolated to
the chiral limit and the corresponding $s\bar{s}$ state from the connected insertion
(no annihilation) is $1.46 \pm 0.05$ GeV. These lattice results are
consistent with the experimental fact that $K_0^{*}(1430)$ is basically
degenerate with $a_0(1450)$ despite having one strange quark. This
resolves one of the puzzles outlined in Sec.~\ref{intro} which is hard
for quark models to accommodate.

   Latest large scale calculation of glueball masses on anisotropic lattices gives
the scalar glueball mass at $1710(50)(80)$ MeV in the quenched approximation~\cite{cad06}
which seems to coincide with $f_0(1700)$ discovered in $J/\Psi$ radiative decays,
long suggested to be a channel for copious glueball production.

\section{Mixing and Decays}

  To answer such questions as raised in Sec.~\ref{intro}, there have been a number of 
studies on the mixing of the isoscalar mesons $f_0(1370), f_0(1500)$, and $f_0(1710)$ 
to sort out their glueball and flavor content of $q\bar{q}$~\cite{ac95}. In considering the
mixing matrix, the usual premise is to place the unmixed (connected insertion without
annihilation) $s\bar{s}$ $\sim 200$ MeV above $n\bar{n}\equiv (u\bar{u}+d\bar{d})/\sqrt{2}$
to reflect the pattern well-known in other meson sectors as well as in baryons. However, this is
not appropriate here. It runs counter to the fact that $K_0^*(1430)$ is basically 
degenerate with $a_0(1450)$. In view of the lattice results discussed in Sec.~\ref{lattice}
where one finds that $a_0(1450), K_0^*(1430)$, and the unmixed $s\bar{s}$ are nearly
degenerate, an apparent conclusion is that the scalar $q\bar{q}$ mesons have, to first order,
an unbroken $SU(3)$ octet. As a result, $f_0(1500)$, which is close to
$a_0(1450), K_0^*(1430)$, should be a fairly pure $f_{octet}= (u\bar{u}+ 
d\bar{d} - 2s\bar{s})/\sqrt{6}$ state. A mixing model, which takes the 
degeneracy of unmixed $n\bar{n}$ and $s\bar{s}$ and the quenched prediction of scalar glueball mass
at $\sim 1700$ MeV into account with slight $SU(3)$ breaking, is quite successful in 
delineating the decays into pseudoscalar pairs of the isoscalar mesons as well as various 
decays from $J/\Psi$. The details of the fit and
predictions are given in a previous work~\cite{ccl06a}. We want to point out several     
salient and robust features in the resultant mixing and decay patterns.

\begin{itemize}
\item
      
$f_0(1500)$ is indeed a fairly pure octet ($f_{octet}$) with very little mixing with the singlet 
and the glueball. $f_0(1710)$ and $f_0(1370)$ are dominated by the glueball and the $q\bar{q}$ 
singlet respectively, with $\sim 10\%$ mixing between the two. This is consisent with the
experimental result $\Gamma(J/\psi\to \gamma f_0(1710))\sim 5\, \Gamma(J/\psi\to \gamma f_0(1500))$
~\cite{ait01} which favors $f_0(1710)$ to have larger glueball component~\cite{ccl06a}.

\item
      The ratio $\Gamma(f_0(1500)\rightarrow  K\overline{K})/\Gamma(f_0(1500)
\rightarrow  \pi\pi) = 0.246\pm0.026$ is one of the best experimentally determined decay ratios
for these mesons~\cite{pdt06}. When the mixing with glueball and $SU(3)$ breaking are neglected,
one obtains
\begin{equation} \label{eq:1500decay}
 {\Gamma(f_0(1500)\to K\overline K)\over\Gamma(f_0(1500)\to
 \pi\pi)} = {1\over 3}\left(1+{s\over u/d}\right)^2{p_K\over
 p_\pi},
\end{equation}   
where $p_h$ is the c.m. momentum of the hadron $h$, $u/d$ and
$s$ are the coefficients for the $u\bar{u}/d\bar{d}$ and $s\bar{s}$ components of the 
$f_0(1500)$ wavefunction. If $f_0(1500)$ is a glueball (i.e. a
flavor singlet) or $s\bar{s}$, the ratio will be 0.84 or larger then unity. 
Either one is much larger than the experimental value. On the other hand, if $f_0(1500)$ is 
$f_{octet}$, then the ratio is $0.21$ which is already close to the experimental number. 
This further demonstrates that $f_0(1500)$ is mainly an octet and its decay ratio can be 
well described with a small $SU(3)$ breaking~\cite{ccl06a}.

\item
Because the $n\bar n$ content is more copious than $s\bar s$ in
$f_0(1710)$ in this mixing scheme, the prediction of $\Gamma(J/\psi\to
\omega f_0(1710))/\Gamma(J/\psi\to \phi f_0(1710))=4.1$ is
naturally large and consistent with the observed value of $6.6\pm2.7$. 
This ratio is not easy to accommodate in a picture where the 
$f_0(1710)$ is dominated by $s\bar s$. One may have to rely
on a doubly OZI suppressed process to dominate over the singly OZI suppressed 
process to explain it~\cite{ac95} .

\end{itemize}

 \begin{figure} [tbh]
\vspace*{-0.5in}
  \centerline{%
    \includegraphics[width=11.0cm]{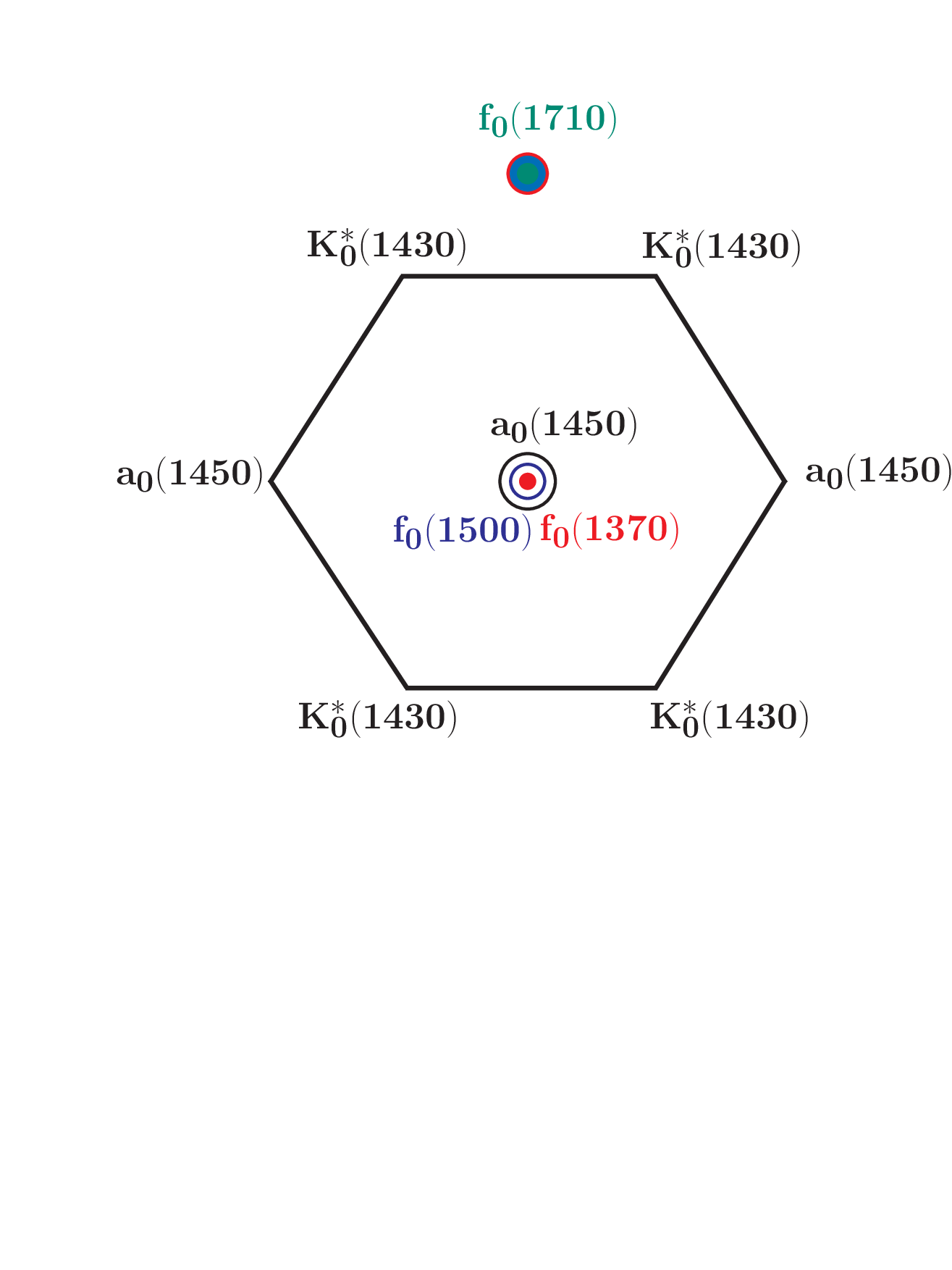} 
 }
  \vspace*{-2in}
\end{figure} 
 \begin{figure} [ht]
\vspace*{-1in}
  \centerline{%
\hspace*{0.3cm}
       \includegraphics[width=11.0cm]{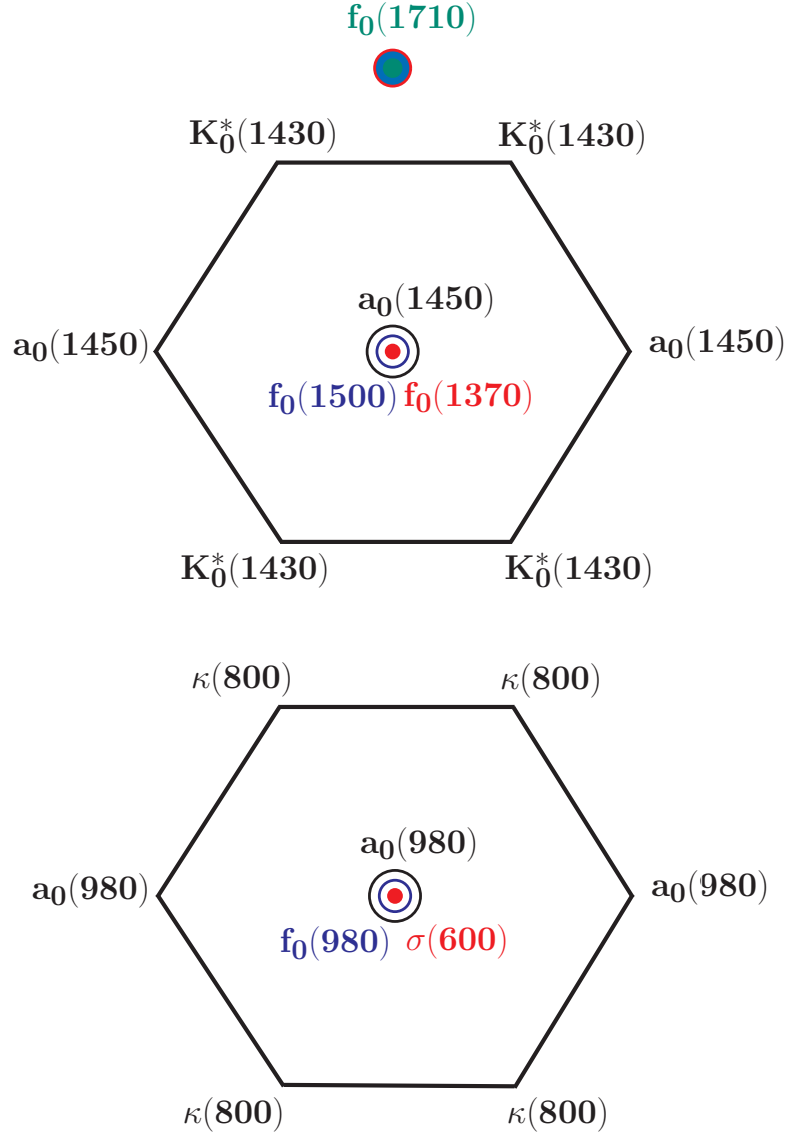}
 }
 \vspace*{-3in}
  \caption{Pattern of light scalar mesons -- a tetraquark mesonium nonet below 1 GeV, 
an almost pure $SU(3)$ $q\bar{q}$ nonet and a nearly pure glueball above 1 GeV.
  }
\end{figure}

\section{Conclusion}

  Notwithstanding many detailed questions remain unanswered satisfactorily,
lattice QCD calculations of scalar meson and glueball masses 
and a phenomenological study of meson decays and their mixing have suggested that 
a pattern for light scalar mesons is starting to arise --
a tetraquark mesonium nonet below 1 GeV, and an almost pure
$SU(3)$ $q\bar{q}$ nonet and a fairly pure glueball above 1 GeV.
It should be scrutinized by experiments in the future, such as with high statistics $J/\Psi$ and
$D$ decays and $p\bar{p}$ annihilation. Lattice calculations with light dynamical
fermions are needed to check the pattern and determine the strong decay widths of these
mesons.

\section*{Acknowledgments}
The author thanks the Yukawa Institute for Theoretical Physics at Kyoto University for its 
hospitality while attending the YKIS2006 workshop on "New Frontiers on QCD". He also thanks 
M. Chanowitz, H.Y. Cheng, C.K. Chua, R. Jaffe, \mbox{T. Kunihiro}, C. Liu, M. Pennington, 
J. Schechter, and Q. Zhao for useful discussions.

%



\begin{thebibliography}{99}

\bibitem{pdt06}
W.-M. Yao {\it et al.}, (Particle Data Group), \JL{J. of Phys.,
33G,2006,1}

\bibitem{ait01}
E.M. Aitala  {\it et al.}, \PRL{86,2001,770}; M. Ablikim  {\it et al.},
\PLB{598,2004,149}; M. Ablikim {\it et al.}, \PRD{72,2005,092002};
M. Ablikim {\it et al.}, \PLB{633,2006,681}; M. Ablikim {\it et al.}, \PLB{642,2006,441}


\bibitem{bar85}
T. Barnes, \PLB{165,1985,434}

\bibitem{Schechter06} 
D. Black, M. Harada and J. Schechter, \PRD{73,2006,054017}

\bibitem{jaf77}
R.L. Jaffe, \PRD{15,1977,267}

\bibitem{lw81}
K.F. Liu and C.W. Wong, \PLB{107,1981,391}

\bibitem{wi82}
J. Weinstein and N. Isgur, \PRL{48,1982,659}

\bibitem{ccl06}
I. Caprin, G. Colangelo and H. Leutwyler, \PRL{96,2006,132001},
[hep-lat/0512346]

\bibitem{ll82}
B.A. Li and K.F. Liu, \PLB{118,1982,435}; \PRD{30,1984,613};
\PRL{51,1983,1510}

\bibitem{ll83}
B.A. Li and K.F. Liu, \PRD{28,1983,1636}; \PRD{29,1984,416};
\PLB{B134,1984,128}

\bibitem{swa06}
See, for example, E.S. Swanson, \PRP{429,2006,243}, [hep-ph/0601110].

\bibitem{mac06}
N. Mathur, A. Alexandru, Y. Chen, S.J. Dong, T. Draper,
Horv\'ath, F.X. Lee, K.F. Liu, S. Tamhankar, and J.B. Zhang,
[hep-ph/0607110]


\bibitem{bg96}
C. Bernard and M. Golterman, \PRD{53,1996,476}, [hep-lat/9507004]

\bibitem{lus86}
M. L\"{u}scher, \CMP{105,1986,153}

\bibitem{rg95}
K. Rummukainen and S. Gottlieb, \NPB{450,1995,397}, [hep-lat/9503028]


\bibitem{lw00}
L.-J. Lee and D. Weingarten, \PRD{61,2000,014015};
M. Gockeler {\it et al}., \PRD{57,1998,5562}; S.
Kim and S. Ohta, \JL{Nucl. Phys. Proc. Suppl.,B53,1997,199};
A. Hart, C. McNeile, and C. Michael, \JL{Nucl. Phys. Proc. Suppl.,
B119,2003,266}; T. Burch {\it et al}., [hep-lat/0601026];
H. Wada {\it et al.}, [hep-lat/0702023]

\bibitem{bde02}
W. Bardeen, A. Duncan, E. Eichten, N. Isgur, and H. Thacker,
\PRD{65,2002,014509}; W. Bardeen, E. Eichten, and H. Thacker,
\PRD{69,2004,054502}

\bibitem{po03}
S. Prelovsek, K. Orginos, \JL{Nucl. Phys. Proc. Suppl.,119,2003,822};
C. Bernard et al., \PRD{64,2001,054506}; 
S. Prelovsek, C. Dawson, T. Izubuchi, K. Orginos, and A. Soni,
\PRD{70,2004,094503}

\bibitem{cad06} Y. Chen, A. Alexandru, S.J. Dong, T. Draper, Horv\'ath,
F.X. Lee, K.F. Liu, N. Mathur, C. Morningstar, M. Peardon, S.
Tamhankar, B.L. Yang, and J.B. Zhang, \PRD{73,2006,014516}

\bibitem{ac95}
C. Amsler and F.E. Close, \PLB{353,1995,385};  \PRD{53,1996,295}; F.E. Close and A. Kirk, 
\PLB{483,2000,345};  F.E. Close and Q. Zhao, \PRD{71,2005,094022}; X.G. He, X.Q. Li, X. Liu, 
and X.Q. Zeng, \PRD{73,2006,051502}; ibid. {73,2006,114026}; W. Lee and D. Weingarten, 
\PRD{61,1999,014015};  F. Giacosa {\it et al.}, \PRD{72,2005,094006};  L. Burakovsky and 
P.R. Page, \PRD{59,1998,014022}

\bibitem{ccl06a}
H.Y. Cheng, C.K. Chua, and K.F. Liu, \PRD{74,2006,094005}




\end{thebibliography}
\end{document}